\documentclass[twocolumn,showpacs,amsmath,amssymb,prl]{revtex4}

\usepackage{graphicx}

\begin{document}


\title{Impurity band in clean superconducting weak links}
\author{N. B. Kopnin $^{(1,2)}$}
\affiliation{$^{(1)}$ Low Temperature Laboratory, Helsinki University of
Technology, P.O. Box 2200, FIN-02015 HUT, Finland,\\
$^{(2)}$ L. D. Landau Institute for Theoretical Physics, 117940 Moscow,
Russia}
\date{\today}

\begin{abstract}
Weak impurity scattering produces a narrow band with a
finite density of states near the phase difference $\phi
=\pi$ in the mid-gap energy spectrum of a macroscopic
superconducting weak link. The equivalent distribution of transmission
coefficients of various conducting quantum channels is found.
\end{abstract}

\pacs{74.80.Fp, 73.23.-b, 73.63.Rt }

\maketitle

Effect of impurities on transport properties of superconducting
nanostructures is one of the key issues in the physics of mesoscopic
superconductors (see \cite{Beenakker/rev} for a review). Recent advances in
nanotechnology revived interest in quantum point contacts (see \cite{Lodder}
and references therein) and in other devices that employ quantum conductors
connected to superconducting electrodes \cite{Pierre}. SNS junctions and
weak links consisting of two superconductors connected by a small orifice in
a thin insulating layer (point contacts) are the simplest devices of
interest. Though impurity effects in SNS junctions are well studied (the
work began with Ref. \cite{Mitsai} and still goes on, see for example \cite%
{Zhou}), the role of impurity scattering in superconducting point contacts
remains not fully investigated.

As an example, consider a ballistic point contact between two 
superconductors assuming
that the thickness $d$ of the insulating layer and the size of the orifice $%
a $ are shorter than the coherence length $\xi$ and the impurity mean free
path $\ell$. As is
well known \cite{Kulik,KulikOmel} there exist mid-gap states with the
spectrum
\begin{equation}
\epsilon =\pm \epsilon _0,\; 
\epsilon _0=|\Delta |\cos (\phi /2)  \label{spectr/clean}
\end{equation}
Here $\phi =\chi _{2}-\chi _{1}$, with $\chi _{1,2}$ being the order parameter 
phase on the left (right) from the orifice; the upper sign refers to 
particles moving to the right from region 1 into region 2 and vice versa. 
Impurities do not
appear in Eq.\ (\ref{spectr/clean}) because the characteristic dimension is
determined by the size of the constriction $d$ rather than by $\xi $, the
parameter $d/\ell $ being assumed infinitely small. It is natural to expect
however that the impurity scattering would modify this spectrum at an energy
scale determined by the small parameter $d/\ell $, especially for low
energies where the two branches of the spectrum for right- and left-moving
particles cross. The same can be expected for 
a long ballistic SNS junction which has a mid-gap spectrum 
\cite{Mitsai,Kulik} consisting of
many branches for right- and left-moving particles that cross at $\phi =\pi$.

In this Letter we consider both point contacts and long SNS junctions
and show that a weak impurity scattering transforms their mid-gap spectra
in such a way that a narrow band having a finite density of 
states and a width $\sim |\Delta|\sqrt{d/\ell}$
appears near each crossing point at the phase difference $%
\phi =\pi$. This impurity band is expected to have a dramatic effect on
dynamic properties of weak links. In particular,
it enhances the inelastic electron-phonon relaxation rate 
at low temperatures which would be otherwise nearly zero
due to the energy conservation. 

\begin{figure}[t]
\centerline{  \includegraphics[width=0.4\linewidth]{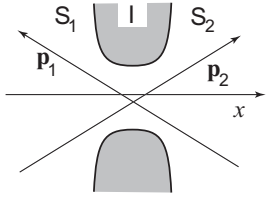}}
 \caption{The point contact between two superconductors ($S_1$ and $S_2$)
 separated by an insulating
 layer (I). The $x$ axis is directed from the region where the order parameter
 phase is $\chi _1$ into the region where the phase is $\chi _2$.}
 \label{pointcontact}
 \end{figure}

{\it Impurity band in a point contact.}--We consider first a ballistic 
point contact such that $a\sim d\ll \xi \ll \ell $. 
One has $\Delta _{1,2}=|\Delta
|e^{i\chi _{1,2}}$ to the left (right) from the orifice, respectively
(see Fig.\ \ref{pointcontact} ). The quasiclassical Green functions 
(retarded or
advanced) satisfy the normalization $g^2-ff^\dagger =1$ and obey the
Eilenberger equations \cite{Eilenberger}
\begin{eqnarray}
-iv_{F}\frac{\partial g}{\partial s} &=&
\frac{i}{2\tau }\left( f\left\langle f^{\dagger }\right\rangle
-\left\langle f\right\rangle f^{\dagger }\right) \label{eileq1}\\
-iv_{F}\frac{\partial f}{\partial s}-2\epsilon f+2\Delta g &=&\frac{i}{\tau }%
\left( \left\langle g\right\rangle f-\left\langle f\right\rangle g\right) ,
\label{eileq2} \\
iv_{F}\frac{\partial f^{\dagger }}{\partial s}-2\epsilon f^{\dagger
}+2\Delta ^{*}g &=&\frac{i}{\tau }\left( \left\langle g\right\rangle
f^{\dagger }-\left\langle f^{\dagger }\right\rangle g\right)  \label{eileq3}
\end{eqnarray}
where $s$ is the distance along the particle trajectory. We assume
zero magnetic field. The right-hand sides of these equations
describe the scattering by impurities. We use $\langle \ldots
\rangle $ to denote an average over the Fermi surface. The
standard technique of averaging over impurities is applicable
because the number of impurities within the volume of the orifice
is large. Indeed, the mean free time is $\tau ^{-1}\sim N_F
n_{{\rm imp}}|u|^{2}$ where $|u|\sim Up_{F}^{-3}$ is the Fourier
transform of the impurity potential $U$, and $N_{F}$ is the normal
single-spin density of states at the Fermi level.
The number of impurities $%
n_{{\rm imp}}da^{2}\sim (d/\ell )(p_{F}a)^{2}(E_{F}/U)^{2}$ can be very
large even for $U\sim E_{F}$ because of a macroscopic number of quantum
channels in the orifice $(p_{F}a)^2\gg 1$.

For low energies $|\epsilon |\ll |\Delta |$ and close to $\phi
=\pi$ the Green functions are localized near the orifice at
distances $s\sim \xi$. We put $\phi =\pi +\delta $ where $\delta
\ll 1$ so that $\Delta _{1} = -i|\Delta |e^{-i\delta /2},\; \Delta
_{2} = i|\Delta |e^{i\delta /2}$. Following \cite{KrPe} we write
\begin{equation}
f_\pm = \mp i\zeta_\pm \pm\eta_\pm ,\; f^{\dagger }_\pm = \mp i\zeta_\pm \mp
\eta_\pm .  \label{etazeta1}
\end{equation}
The upper sign is for right-moving particles, $p_x>0$, the lower sign is for
left-moving particles, $p_x<0$. The $x$ axis is perpendicular to the
insulating layer. We assume $\zeta ^{2}\gg 1-\eta ^{2}$ so that $g_\pm
=i\zeta _\pm$.
Equations (\ref{eileq2}, \ref{eileq3}) become
\begin{eqnarray*}
v_{F}\frac{\partial \zeta _{\pm}}{\partial s}+2\zeta _{\pm}|\Delta | \cos
(\delta /2){\rm sign}(s) =0 , \\
v_{F}\frac{\partial \eta _{\pm}}{\partial s}-\left( 2\epsilon \mp 2|\Delta
|\sin (\delta /2)+i\sigma _\pm \right) \zeta _{\pm} =0,
\end{eqnarray*}
where 
\begin{equation}
\sigma _\pm =\tau ^{-1}\left( \left\langle g\right\rangle \pm
\left\langle f\right\rangle /2 \pm \left\langle f^{\dagger }\right\rangle /2
\right) . \label{sigma}
\end{equation}

The boundary conditions for $s\rightarrow \infty $ follow from the
expressions for the Green functions in the bulk. For small $\delta
$, the functions $\zeta _{\pm }\rightarrow 0$ and $\eta _{\pm
}\rightarrow 1$. The solution is
\begin{eqnarray*}
\zeta _{\pm } &=&C_{\pm }\exp \left( -K\right) ,\;K=2|\Delta ||s|/v_{F}, \\
\eta _{\pm } &=&C_{\pm }v_F^{-1}\int_{0}^{s}e^{-K} \left( 2|\Delta|E_\pm
+i\sigma _\pm \right) \,ds ,\\
C_{\pm }&=&\left[E_\pm +iv_F^{-1}
\int_{0}^{\infty }\sigma _\pm \,ds\right] ^{-1} .
\end{eqnarray*}
Here $E_{\pm }=\epsilon /|\Delta |\mp \sin (\delta /2)$.

The averages $\left\langle g\right\rangle $, etc., are
proportional to the solid angle at which the orifice is visible
from the position point, they decrease quickly at distances $s\sim
a$ from the orifice. Indeed, $g\left( {\bf p},{\bf r}\right) =\mp
f\left( {\bf p},{\bf r}\right) =i\zeta \left(
0\right) $ for trajectories that go through the orifice and $g\left( {\bf p},%
{\bf r}\right) =f\left( {\bf p},{\bf r}\right) =0$ otherwise
because, for non-through trajectories, the Green functions are
small as compared to $\zeta \gg 1$. For $s\sim a$ one has $K=0$.
Moreover, using Eq.\ (\ref{etazeta1}) we find
\[
v_F^{-1} \int \sigma _\pm\, ds =\ell ^{-1}\int ds \left\langle g(0)
\right\rangle _\mp =\gamma g_\mp (0)
\]
where $\left\langle g \right\rangle _\pm =(2\pi )^{-1} \int_{\pm p_{x}>0}g(%
{\bf p},{\bf r})\, d\Omega _{{\bf p}}$, and $\gamma \sim d/\ell $ is a
geometric factor depending on the shape of the contact and on the position
of the trajectory with respect to the orifice. We neglect the latter
dependence and consider a constant $\gamma $ for simplicity. Finally,
\begin{equation}
g_{\pm}^{R(A)}\left( s\right) =ie^{-K}\left[E_{\pm} +i\gamma
g_{\mp}^{R(A)}\left( 0\right) \right]^{-1}.  \label{g-inside1}
\end{equation}

\begin{figure}[t!!!]
\centerline{  \includegraphics[width=0.6\linewidth]{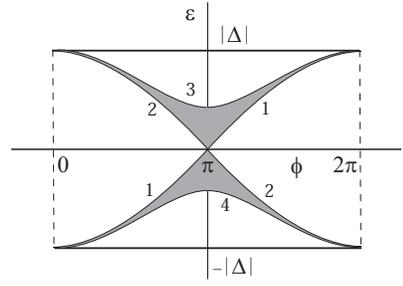}}
\medskip
\caption{ Lines 1, 2, 3, and 4 encompass
the energy band with a finite density of states (shaded region).
The maximum half-width of the band is $2\sqrt{\gamma}$. For a
ballistic contact $\tau \rightarrow \infty$, the spectrum is given
by Eq.\ (\protect\ref{spectr/clean}) and follows lines 1 and 2. 
The entire pattern is $2\pi$-periodic in $\phi$. } \label{fig-spectr}
\end{figure}

In the limit $\gamma \rightarrow 0$, the Green functions have poles at $%
E_\pm =0$ which is Eq.\ (\ref{spectr/clean}). The spectrum is
shown in Fig. \ref{fig-spectr} by lines (1) for $p_{x}>0$ and (2)
for $p_{x}<0$. However, as $E_{+}E_{-}$ approaches $\gamma $, the
apparent pole-like behavior of Eq.\ (\ref{g-inside1}) transforms
into a more complicated dependence. To calculate the Green
functions, we solve Eq.\ (\ref {g-inside1}) for $g_{\pm }(0)$. If
$0<E_{+}E_{-}<4\gamma $, i.e.,
\begin{equation}
\sin ^{2}(\delta /2)<\epsilon ^{2}/|\Delta |^{2}<4\gamma +\sin ^{2}(\delta
/2)  \label{e-range}
\end{equation}
we find $g_{\pm }^{R}\left( s\right) =g_{\pm }^{R}\left( 0\right) e^{-K}$
where
\begin{equation}
g_{\pm }^{R}\left( 0\right) =\frac{E_{\mp }}{2\gamma }\left( i+\sqrt{\frac{%
4\gamma }{E_{+}E_{-}}-1}\right) ,  \label{g-inside3}
\end{equation}
and $g^{A}=-[g^{R}]^{*}$. The radical is defined as an analytical function
with the cuts along the borders of the region
determined by Eq.\ (\ref{e-range}) (shaded region in Fig.\ \ref{fig-spectr}). 
The sign is chosen for $\epsilon $ within the upper part 
of the region. The normalized density of states $\nu
_{\epsilon }(s)=\left[ g^{R}\left( s\right) -g^{A}\left( s\right)
\right] /2$ is nonzero within the energy interval of Eq.\
(\ref{e-range}). The maximum half-width of the energy band $2\sqrt{\gamma
}$ is reached for $\delta =0$. Inside the constriction
\begin{equation}
\nu _{\epsilon \pm }(0)=\frac{E_{\mp }}{2\gamma }\sqrt{\frac{4\gamma }{%
E_{+}E_{-}}-1}  \label{dos}
\end{equation}
For a particle with a given sign of $p_x$ the density of states is nonzero
near both $E_- =0$ and $E_+=0$: the  scattering mixes
states with positive and negative $p_x$. However, far from the crossing
point, $|\epsilon |\gg |\Delta|\sqrt{\gamma}$, the ratio 
$\nu _\pm/\nu _\mp =E_\mp/E_\pm$ is large near the corresponding ballistic 
spectrum $E_\pm =0$: its magnitude is of the
order $\epsilon ^2/|\Delta |^2\gamma$.

Beyond the range of Eq.\ (\ref{e-range}) $g^{R}$ and $g^{A}$
coincide and the density of states vanishes. For $E_+E_->4\gamma$,
\[
g_{\pm }^{R}(0)=g_{\pm }^{A}(0)=\frac{iE_{\mp }}{2\gamma }\left( 1-\sqrt{1-%
\frac{4\gamma }{E_{+}E_{-}}}\right) .
\]
For $\gamma \ll |E_{+}E_{-}|$ one recovers the poles $g_{\pm }^{R(A)}\left(
0\right) =i/E_{\pm }$ with the density of states $\nu = \pi |\Delta|
\delta (\epsilon \mp \epsilon _0)$. 
Obviously, Eq.\ (\ref{dos}) conserves the
total number of states since
$\int \nu _\pm \, d\epsilon =\pi |\Delta|$ for any $\phi$.

\begin{figure}[t!!!]
 \centerline{  \includegraphics[width=0.6\linewidth]{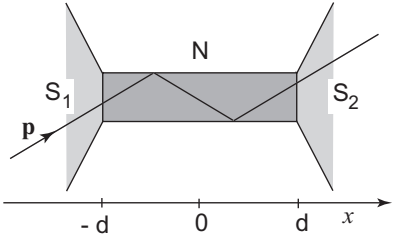}}
 \caption{Normal bridge between two bulk superconductors.}
 \label{fig-bridge}
 \end{figure}

{\it Long SNS bridge.}--
Consider now a normal bridge of a length $d\gg \xi$  that connects two
massive superconductors. Its width is much
shorter than $\xi$. We assume that it has specular walls. Irregularities on
the walls can also be
modelled by random impurities. Superconductors and the normal
metal have the same Fermi velocities $v_{F}$. We take the $x$ axis along the
bridge (see Fig. \ref{fig-bridge}).

Let $s=\pm s_0=\pm d/|\cos \theta |$ be the
outlets of the normal bridge into the bulk superconductors.
For $s<-s_0$ the solutions in the bulk are \cite{Mitsai,Kulik} 
\begin{equation}
g_\pm =g_{\infty }+g_{\pm k}e^{k(s+s_0)},~ 
f_\pm =f_{\infty }+g_{\pm k}\beta _{k}e^{k(s+s_0)}
\label{g-bulk}
\end{equation}
where $f_\infty =\Delta /i\sqrt{|\Delta |^2-\epsilon ^2}$,
$g_\infty =\epsilon /i\sqrt{|\Delta |^2-\epsilon ^2}$ while
$k=2v_F^{-1} \sqrt{\left| \Delta \right| ^{2}-\epsilon ^{2}}$ and
$\beta _k =\Delta/\left(\epsilon +iv_Fk/2\right)$. The
impurity scattering in the bulk can be neglected.
The signs correspond to $\pm p_x>0$; the order parameter 
phase should be $\chi _1$ for $p_x >0$ and $\chi _2$ for $p_x<0$.
For $s>s_0$
\begin{equation}
g_\pm =g_{\infty }+g_{\pm k}e^{-k(s-s_0)},~ 
f_\pm =f_{\infty }+g_{\pm k}\beta _{-k}e^{-k(s-s_0)} .
\label{g-bulk1}
\end{equation}
Here the phase is $\chi _2$ for $p_x>0$ and $\chi _1$ for $p_x<0$.

\begin{figure}[t!!!]
\centerline{  \includegraphics[width=0.55\linewidth]{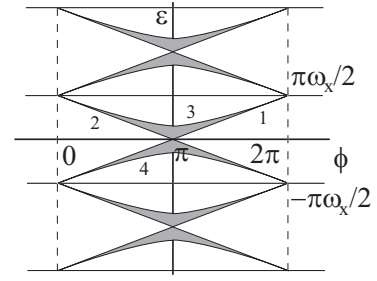}}
\medskip
\caption{The energy bands (shaded regions) for a long SNS bridge. Lines 1 and 2
show the spectrum of a particle with $p_x>0$ and $p_x<0$, respectively,
in a ballistic bridge $\ell \rightarrow \infty$. The pattern 
is $2\pi$-periodic in $\phi$.} 
\label{snsspec2}
\end{figure}

We first outline the known solution \cite{Mitsai,Kulik} 
for states with not very low energies 
$v_F/d\ll \epsilon $ such that $d\gg \xi _{N}$
Here $\xi _{N}\sim v_{F}/T$. In the region inside the bridge where $\Delta =0$
Eqs.\ (\ref{eileq1}--\ref{eileq3}) yield
$g^2\equiv g^2_{N}=const$ and
\[
f= C\exp \left[ i\left( 2\epsilon +\frac{i}{\tau }\left\langle
g_{N}\right\rangle \right) \frac{x}{v_{x}}\right] 
\]
where $v_x=v_F\cos \theta$ and $x=s\cos \theta$.  We assumed
that $\left\langle f\right\rangle =\left\langle f^{\dagger
}\right\rangle =0$. Indeed, $f$ and $f^{\dagger }$
oscillate rapidly as functions of $\theta $ since $\epsilon
d/v_{F}\gg 1$ and vanish after averaging over the angles.  
The continuity at the borders between the bridge and the bulk gives
$g_{\pm k}=g_{\pm N}-g_{\infty }$
and
\begin{equation}
g_{\pm N} = \frac{\tan \alpha \pm ig_{\infty }}{g_{\infty }\tan \alpha \pm i}
\label{g-N}
\end{equation}
where 
$\alpha =(d/v_x)\left( 2\epsilon +i\left\langle g_{N}\right\rangle /\tau
\right)- \phi /2$. For $\left| \epsilon \right|
<\left| \Delta \right| $, the function $g_{\pm N}$ has poles  
when $\alpha =\pm \arccos \left( \epsilon /\left|
\Delta \right| \right) +\pi n$ or \cite{Mitsai,Kulik}
\[
\epsilon +\frac{i}{2\tau} \left\langle g_N\right\rangle =\pm
\omega _x\left[ \frac{\phi}{2}+ \pi
\left(n+\frac{1}{2}\right)-\arcsin\left(\frac{\epsilon}{|\Delta
|}\right)\right]
\]
where $\omega _x=|v_x|/2d$.
The state with $n=-1$ has $\epsilon =0$ for $\phi =\pi$. The
spectrum for $\tau =\infty$ is shown in Fig. \ref{snsspec2}
by lines 1 and 2. For $\epsilon \gg v_F/d$, the poles are closely packed. 
As a result, the angular average
collects contributions from many poles which gives\cite{Mitsai}
$\left\langle g_{N}^{R\left( A\right) }\right\rangle =\pm 1$.

For low energies $\epsilon \ll |\Delta|$, the term 
$g_\infty \tan \alpha \ll 1$ if  $d\gg \xi$. Eq.\ (\ref{g-N})
results in
$g_N= \tan \alpha$. The Green function is large
when $\alpha$ is close to $\pi /2 +\pi n$, i.e., close to the points
$\epsilon =\omega _0 \pi m$ where the branches for $p_x>0$ and for
$p_x<0$ cross. Consider these regions in more detail.
We put $\epsilon =\epsilon ^\prime +\omega _0 \pi m$ and write
the functions in the normal bridge as $f=\tilde f \exp (\pm i\pi mx/d)$, 
$f^\dagger ={\tilde f}^\dagger \exp (\mp i\pi mx/d)$. We now express
$\tilde f$ and ${\tilde f}^\dagger$ through 
$\eta$ and $\zeta$ according to Eq. (\ref{etazeta1}) and assume
$g _N =i\zeta$.
The solution of Eqs. (\ref{eileq2}, \ref{eileq3}) is
$\zeta _\pm = A _\pm =const $ and
\[
\eta _\pm = A_\pm v_x^{-1}\int_{0}^{x}\left( 2\epsilon ^\prime +i\sigma _\pm
 \right) \, dx .
\]
Here $\sigma _\pm$ is determined by Eq.\ (\ref{sigma}) where $f,\, f^\dagger$
are replaced with $\tilde f,\, {\tilde f}^\dagger$.
Note that now $K=0$. Matching this with Eqs.\ (\ref{g-bulk}, \ref{g-bulk1}) 
at $x =\pm d$  we obtain
\begin{equation}
g_{\pm N}= \frac{i\omega _x}{\epsilon^\prime +i\sigma_\pm /2 
\mp \omega _x \delta /2}  \label{gN-final}
\end{equation}
where $\phi =\pi +\delta $. Using Eq. (\ref{etazeta1}) we find $\sigma _\pm =
\left\langle g\right\rangle _\mp /\tau$.

The angle-resolved density of states is proportional to 
the averaged $g^R-g^A$. Calculating the average we obtain
\begin{equation}
\left\langle g_N \right\rangle_\pm = i\omega _0
\int _0^1 \frac{\mu\, d\mu }{\epsilon^\prime 
+i\left\langle g \right\rangle_\mp /2\tau
\mp \mu \omega _0\delta /2} \label{g-intexp}
\end{equation}
where $\omega _0=v_F/2d$. We put
$ x=\delta/2\sqrt{\gamma},\;
y=\epsilon ^\prime /\omega _0\sqrt{\gamma},
\;
z_\pm =\sqrt{\gamma}\left\langle g_N \right\rangle_\pm$
where $\gamma =d/\ell$. 
Performing integration in Eq.\ (\ref{g-intexp}) 
for $\delta \sim \sqrt{\gamma}$ and $\epsilon \sim \omega _0\sqrt{\gamma}$
we find
\begin{equation}
z_\pm = \mp \frac{i}{x} -\frac{i(y+iz_\mp)}{x^2}
\ln \left[ \frac{y+iz_\mp \mp x}{y+iz_\mp}\right] .
\end{equation}

For small $x$ one has $z_+=z_-=\left[ iy \pm \sqrt{2 -y^2} \right]/2$
which is similar Eq. (\ref{g-inside3}).
For $y<\sqrt{2}$, the functions $\left\langle g^R\right\rangle
=-\left\langle g^A\right\rangle ^*$ are complex, thus the density of states
is nonzero, $\left\langle\nu\right\rangle 
\sim 1/\sqrt{\gamma}$.
For $y>\sqrt{2}$ solutions are imaginary, $\nu =0$.
To find the limit where
a nonzero density of states appears, we look for $z =ia$ and find conditions
when there are no real solutions for $a$. We have
\begin{equation}
a_\pm =\mp \frac{1}{x} -\frac{y-a_\mp}{x^2}\ln \left[
\frac{y-a_\mp \mp x}{y-a_\mp}\right] . \label{border-eq}
\end{equation}
For small $x$ we return to the previous result: the density of states 
vanishes for $y>\sqrt{2}$. However, if $\epsilon^\prime =y =0$ a real
solution for $a$ exists for any
$x$: the density of states is zero along the axis
$\epsilon^\prime =0$ which agrees with Eq. (\ref{e-range}). Next, we note
that the density of states is singular at one of the borders of
its region of existence. Thus, either $a_+$ or $a_-$ should go to infinity.
Assume that $a_+\rightarrow \infty$. Then $a_-\rightarrow 0$ according to
the second equation (\ref{border-eq}). In turn, 
\[
a_+=-\frac{1}{x} -\frac{y}{x^2}\ln \left[ \frac{y-x}{y}\right].
\]
Infinite values for $a_+$ are possible for $y=x$. 
Similarly, assuming $a_-\rightarrow \infty ,\;
a_+\rightarrow 0$ we finally find
the borders of the region with a nonzero density of states, $y=\pm x$ or 
\begin{equation}
|\epsilon ^\prime| =\omega _0|\delta |/2 . \label{innerborder}
\end{equation}
Equation (\ref{innerborder}) gives the
lower-$|\epsilon ^\prime|$ limits shown in Fig.\
\ref{snsspec2} by lines 1 and 2.
The higher-energy limits (lines 3 and 4 in Fig.\ \ref{snsspec2}) 
are set by $|y|=\sqrt{2}$ for small $x$. For large
$x$ and $y$ we expect that one of $a_\pm$ is large while
the other is small. We then again obtain Eq. (\ref{innerborder})
as the asymptotic expression. Therefore, the two borders approach
each other for large $x$ and $y$, as was also the case for a point
contact. We thus expect that Eq.\ (\ref{dos}) provides a
qualitatively correct representation for the averaged density of states
of a long SNS junction, as well.

{\it Discussion.}--The results for point contacts and for long SNS bridges 
are qualitatively similar. To simplify the discussion we consider the 
impurity band in
a point contact where an explicit equation for the density of states is 
available. It can be compared with the known mid-gap 
energy spectrum\cite{Zaitsev,Beenakker} for a contact that has a 
tunnel barrier with a transmission probability $D$,
\begin{equation}
\epsilon =\pm \epsilon _D,\;
\epsilon_D = |\Delta |\sqrt{1-D\sin ^2(\phi /2)}.  \label{spectr/barrier}
\end{equation}
In Fig. \ref{fig-spectr} this spectrum follows lines (3) and (4) and has a
gap with the half-width $|\Delta |\sqrt{R}$ at $\phi =\pi$ where $%
R=1-D$ is the reflection coefficient. The gap results from the mixing of
right- and left-moving particles provided by the barrier. 
One can consider the impurity band Eq.\ (\ref{e-range}) as a result of
superposition of a large number of conduction channels with various
transmission coefficients $D$ resulting from scattering of particles on 
different
trajectories characterized by a given momentum. Superposition of 
states with different momenta
due to scattering by impurities is also known to produce an impurity band in
$d$-wave superconductors near the gap nodes \cite{GorkKal}.
To find the equivalent probability distribution $P(D)$ we write 
the total density of states for an energy $\epsilon$ in the form
\[
\left\langle \nu \right\rangle _+ +
\left\langle \nu \right\rangle _- =
\pi |\Delta| \int \left[ \delta \left(\epsilon -\epsilon _D\right)
+\delta \left(\epsilon +\epsilon _D\right)\right] P(D)\, dD .
\]
The relevant values of $D$ are close to one. With help of Eqs.\ (\ref{dos})
and (\ref{spectr/barrier}) we find
\begin{equation}
P(D)=\frac{1}{2\pi \gamma}\, \frac{\sqrt{4\gamma -R}}{\sqrt{R}} ,\;
R<4\gamma . \label{distribution}
\end{equation}
The distribution is truncated at $R\geq 4\gamma$. For $\gamma \rightarrow 0$ 
one obtains $P(D)=\delta (R)$. The square-root singularity
at $R=0$ in Eq.\ (\ref{distribution}) resembles that of
the universal distribution \cite{Dorokhov}
for diffusive normal conductors. The singularity  ensures a finite
density of states at $\phi =\pi$ which is $\nu = (1/\sqrt{\gamma})
\sqrt{1-(\epsilon^2/4\gamma |\Delta |^2)}$ according to Eq.\ (\ref{dos}).
However, Eq.\ (\ref{distribution}) is not 
universal: in contrast to its diffusive counterpart it depends on the
geometry of the contact and on the quasiparticle mean free path.

The impurity band is expected to have a profound
effect on dynamics of weak links. For example, it enhances
the electron-phonon relaxation at low temperatures which otherwise would 
almost vanish due to the energy conservation. Indeed, the electron-phonon
relaxation rate is proportional to the product of two electron densities of 
states $\nu _\epsilon $ and $\nu _{\epsilon _1} $, and the phonon density of 
states $N_{\rm ph}(\omega )$ where $\omega =\epsilon -\epsilon _1$.
Since $N_{\rm ph}(\omega )\propto \omega ^2$ the product vanishes if
each electronic density of states is a delta function $\delta (\epsilon -
\epsilon _0)$ with only one state $\epsilon _0$ for a given phase $\phi$
and the sign of momentum $p_x$. It is the impurity scattering that 
broadens the density of
states into a band thus making the electron-phonon relaxation possible.

I thank D. Averin, Yu. Barash, M. Feigel'man, and G. Volovik for instructive
discussions. This work was supported by Russian Foundation for
Basic Research.


\end{document}